\documentclass[prb,superscriptaddress,twocolumn,floatfix,amsmath,amssymb,showpacs]{revtex4}

\usepackage{graphicx}
\usepackage{dcolumn}
\usepackage{bm}

\begin{document}

\title{Superconductivity, charge- or spin-density wave, and metal-nonmetal transition in BaTi$_{2}$(Sb$_{1-x}$Bi$_{x}$)$_{2}$O}

\author{Hui-Fei Zhai}
\affiliation{Department of Physics, State Key Lab of Silicon
Materials, and Center for Correlated Matter, Zhejiang University,
Hangzhou 310027, China}

\author{Wen-He Jiao}
\affiliation{Department of Physics, State Key Lab of Silicon
Materials, and Center for Correlated Matter, Zhejiang University,
Hangzhou 310027, China}

\author{Yun-Lei Sun}
\affiliation{Department of Physics, State Key Lab of Silicon
Materials, and Center for Correlated Matter, Zhejiang University,
Hangzhou 310027, China}

\author{Jin-Ke Bao}
\affiliation{Department of Physics, State Key Lab of Silicon
Materials, and Center for Correlated Matter, Zhejiang University,
Hangzhou 310027, China}

\author{Hao Jiang}
\affiliation{Department of Physics, State Key Lab of Silicon
Materials, and Center for Correlated Matter, Zhejiang University,
Hangzhou 310027, China}

\author{Xiao-Jun Yang}
\affiliation{Department of Physics, State Key Lab of Silicon
Materials, and Center for Correlated Matter, Zhejiang University,
Hangzhou 310027, China}

\author{Zhang-Tu Tang}
\affiliation{Department of Physics, State Key Lab of Silicon
Materials, and Center for Correlated Matter, Zhejiang University,
Hangzhou 310027, China}

\author{Qian Tao}
\affiliation{Department of Physics, State Key Lab of Silicon
Materials, and Center for Correlated Matter, Zhejiang University,
Hangzhou 310027, China}

\author{Xiao-Feng Xu}
\affiliation{Condensed Matter Physics Group,
  Department of Physics, Hangzhou Normal University, Hangzhou 310036, China}

\author{Yu-Ke Li}
\affiliation{Condensed Matter Physics Group,
  Department of Physics, Hangzhou Normal University, Hangzhou 310036, China}

\author{Chao Cao}
\affiliation{Condensed Matter Physics Group,
  Department of Physics, Hangzhou Normal University, Hangzhou 310036, China}

\author{Jian-Hui Dai}
\affiliation{Condensed Matter Physics Group,
  Department of Physics, Hangzhou Normal University, Hangzhou 310036, China}

\author{Zhu-An Xu}
\affiliation{Department of Physics, State Key Lab of Silicon
Materials, and Center for Correlated Matter, Zhejiang University,
Hangzhou 310027, China}

\author{Guang-Han Cao}
\email[E-mail address: ]{ghcao@zju.edu.cn} \affiliation{Department
of Physics, State Key Lab of Silicon Materials, and Center for
Correlated Matter, Zhejiang University, Hangzhou 310027, China}

\begin{abstract}
We have performed an isovalent substitution study in a layered
titanium oxypnictide system BaTi$_{2}$(Sb$_{1-x}$Bi$_{x}$)$_{2}$O
(0$\leq x\leq$ 0.40) by the measurements of x-ray diffraction,
electrical resistivity and magnetic susceptibility. The parent
compound BaTi$_{2}$Sb$_{2}$O is confirmed to exhibit
superconductivity at 1.5 K as well as charge- or spin-density wave
(CDW/SDW) ordering below 55 K. With the partial substitution of Sb
by Bi, the lattice parameters $a$, $c$ and $c/a$ all increase
monotonically, indicating a negative chemical pressure and lattice
distortion for the (super)conducting Ti$_2$Sb$_2$O-layers. The Bi
doping elevates the superconducting transition temperature to its
maximum $T_c$=3.7 K at $x=$0.17, and then $T_c$ decreases gradually
with further Bi doping. A metal-to-nonmetal transition takes place
around $x$=0.3, and superconductivity at $\sim$1 K survives at the
nonmetal side. The CDW/SDW anomaly, in comparison, is rapidly
suppressed by the Bi doping, and vanishes for $x\geq$0.17. The
results are discussed in terms of negative chemical pressure and
disorder effect.
\end{abstract}



\maketitle

Superconductivity (SC) and charge- or spin-density wave (CDW/SDW, or
abbreviated as DW) are different collective electronic phenomena in
crystalline materials. The DW state often appears in low dimensional
metallic system in which the Fermi surfaces (FSs) are nested,
showing a real-space modulation of charges or spins. SC can be
regarded as another kind of FS instability due to Cooper pairing,
which exhibits an electronic ordering in momentum space in the form
of condensation of Cooper pairs. While SC generally competes with
DW, coexistence of SC and DW is possible.\cite{review} In iron-based
superconducting systems, SC emerges from,\cite{Fe-sdw} or may also
coexist with,\cite{cxhepl} an antiferromagnetic SDW state, which
arouse enormous and intensive researches.\cite{Fe-review}

A possible CDW/SDW anomaly has been observed in a class of titanium
oxypnictides since
1990s.\cite{adam1990,axtell,ozawa2001,ozawa-review,cxh2221,cxh1221,cxh22221}
The material contains Ti$_{2}$O square lattice that was considered
to play an important role for the
anomaly.\cite{pickett,TBA,spin-liquid} The Ti$_{2}$O-sheets can be
viewed as an analogue of CuO$_2$-planes in cuprate superconductors
(if Cu and O atoms of the latter are replaced by O and Ti,
respectively, Ti$_{2}$O lattice forms). Besides, the Ti valence is
3+, giving a $d^1$ configuration for Ti$^{3+}$, in contrast with the
$d^9$ configuration for Cu$^{2+}$ in cuprates. Therefore, continuous
efforts have been made to explore possible SC in these layered
titanium oxypnictides.\cite{adam1990,ozawa-review,cxh1221,cxh2221}
But it was not until very recently that SC was observed in the
related systems. Sun \emph{et al.}\cite{sun} observed SC at 21 K and
DW anomaly at 125 K in an intergrowth compound
Ba$_2$Ti$_2$Fe$_2$As$_4$O containing both Fe$_2$As$_2$ and Ti$_{2}$O
layers. However, the SC was believed to stem from the
Fe$_2$As$_2$-layers rather than the Ti$_{2}$O-sheets. Very recently,
Yajima \emph{et al.}\cite{yajima} reported SC at 1.2 K as well as a
DW transition at 50 K in BaTi$_{2}$Sb$_{2}$O. Doan \emph{et
al.}\cite{doan} found that, by the hole doping with sodium, the
superconducting transition temperature, $T_c$, increased to 5.5 K.
Meanwhile the DW transition temperature, $T_{\text{DW}}$, decreased
to about 30 K. The result suggests competing interplay between SC
and the DW ordering.

The new findings call for investigation of the nature of DW state
and its relation to SC. Earlier neutron diffraction study of
Na$_{2}$Ti$_{2}$Sb$_{2}$O\cite{ozawa2000} failed to observe any
long-range magnetic ordering associated with the resistivity anomaly
at 120 K. Instead, only a structural distortion in the
Ti$_2$Sb$_2$O-layer was found. On the other hand, theoretical
calculations\cite{pickett,TBA} suggest nearly two-dimensional FS
nesting that points to a DW instability. Recent first-principles
calculations tend to favor SDW scenarios in
BaTi$_{2}$As$_{2}$O,\cite{jiang}
Ba$_2$Ti$_2$Fe$_2$As$_4$O,\cite{jiang}
BaTi$_{2}$Sb$_{2}$O,\cite{singh} Na$_2$Ti$_{2}$As$_{2}$O,\cite{yan}
and Na$_2$Ti$_{2}$Sb$_{2}$O\cite{yan}. If the SC is in proximity to
an SDW phase, unconventional SC mediated by spin fluctuations could
be expected.\cite{singh} Nevertheless, CDW instability was also
supported by the calculations of phonon dispersions and
electron-phonon coupling for BaTi$_{2}$Sb$_{2}$O.\cite{subedi}

Previous reports\cite{ozawa2001,cxh2221,cxh22221,yajima,doan}
suggest that $T_{\text{DW}}$ decreases remarkably when As$^{3-}$ is
replaced by Sb$^{3-}$ accompanying with a lattice expansion. SC
appears as DW is sufficiently suppressed.\cite{yajima,doan}
Therefore, it is of great interest to investigate the effect of
isovalent substitution by the last pnictogen Bi in
BaTi$_{2}$Sb$_{2}$O.\cite{note1} In this paper, we present a
systematic Bi-substitution study in the
BaTi$_{2}$(Sb$_{1-x}$Bi$_{x}$)$_{2}$O (1221) system. As expected,
the lattice is expanded by the partial substitution. $T_c$ increases
up to 3.7 K at $x$=0.17, and concomitantly the DW anomaly is
suppressed. Surprisingly, a metal-to-nonmetal transition takes place
around $x$=0.3 where SC is still robust. The results suggest that
the negative chemical pressure suppresses DW and enhances SC, and
the concomitant disorder tends to destroy both DW and SC in
BaTi$_{2}$(Sb$_{1-x}$Bi$_{x}$)$_{2}$O.

Polycrystalline samples of BaTi$_{2}$(Sb$_{1-x'}$Bi$_{x'}$)$_{2}$O
with nominal Bi content $x'$=0, 0.05, 0.1, 0.15, 0.2, 0.25, 0.3,
0.35, 0.4, 0.45 and 0.5 were synthesized by solid state reaction in
vacuum using the staring material of powders of BaO (Alfa Aesar,
99.5\%), Ti (Alfa Aesar, 99.9\%), Sb (Alfa Aesar, 99.5\%) and Bi
(Alfa Aesar, 99.5\%). To obtain a dense pellet that favors the
subsequent resistance measurement, intermediate products of
"Ti$_2$Sb$_3$" and TiBi were prepared respectively at 923 K in an
evacuated quartz tube for 24h. Then the stoichiometric mixtures of
BaTi$_{2}$(Sb$_{1-x'}$Bi$_{x'}$)$_{2}$O were ground in an agate
mortar, and pressed into pellets under a pressure of 2000
kg/cm$^{2}$, in a glove box filled with pure argon (the water and
oxygen content was below 0.1 ppm). The pellets, wrapped with Ta
foils, were sintered at 1323 K for 30h in a sealed evacuated quartz
ampoule, followed by naturally cooling to room temperature. The
as-prepared pellets were very sensitive to moist air. Exposure in
ambient conditions for a few hours led to decomposition of the 1221
phase completely.

Powder x-ray diffraction (XRD) was carried out at room temperature
using a PANalytical x-ray diffractometer (Model EMPYREAN) with a
monochromatic CuK$_{\alpha1}$ radiation. The lattice parameters were
obtained by least-squares fit of more than 20 XRD reflections with
the correction of zero shift, using space group of
\emph{P}4/\emph{mmm} as previously
proposed.\cite{cxh1221,yajima,doan} Energy dispersive x-ray
spectroscopy (EDXS) on a single crystalline grain under a
field-emission scanning electron microscope was used to determine
the exact composition, especially for the incorporated Bi content.
The measurement precision was within $\pm$5\% for the elements Ba,
Ti, Sb and Bi.

Temperature-dependent resistivity was measured in a Cryogenic
Mini-CFM measurement system by a standard four-terminal method.
Additional resistivity measurements down to 0.5 K were carried out
in a $^3$He refrigerator inserted in a Quantum Design PPMS-9
instrument. Gold wires were attached onto the samples' newly-abraded
surface with silver paint, keeping least exposure in air. The size
of the contact pads leads to a total uncertainty in the absolute
values of resistivity of $\pm$15 \%. Temperature-dependent dc
magnetic susceptibility was performed on a Quantum Design MPMS-5
equipment. Both the zero-field-cooling (ZFC) and field-cooling (FC)
protocols were employed under the field of 10 Oe for probing
superconducting transitions. A magnetic field of $H$=10 kOe was
applied for tracking the DW anomaly.

Samples of BaTi$_{2}$(Sb$_{1-x}$Bi$_{x}$)$_{2}$O were first
characterized by powder XRD. Most of the XRD reflections can be well
indexed by a tetragonal lattice, as depicted in the inset of figure
1(a), with $a\sim$4.11 {\AA} and $c\sim$8.10-8.20 {\AA}. Tiny metal
Bi was segregated from the main 1221 phase for low-doping
($x'\leq$0.2) samples, and the amount of Bi and BaTiO$_3$ impurities
increases a little for the high-doping samples. With the Bi doping,
the XRD peaks shift systematically[figure 1(b)], suggesting that
most bismuth incorporates the lattice. The actual Bi content ($x$)
in the 1221 phase was determined by EDXS which showed nearly 20\%
less than the nominal value $x'$, depending on the doping levels and
synthetic conditions. The result gives $x$=0, 0.04, 0.08, 0.13,
0.17, 0.20, 0.25, 0.3, 0.35, and 0.40 for the nominal $x'$ values of
0, 0.05, 0.1, 0.15, 0.2, 0.25, 0.3, 0.35, 0.4, and 0.45,
respectively.

\begin{figure}
\includegraphics[width=7.5cm]{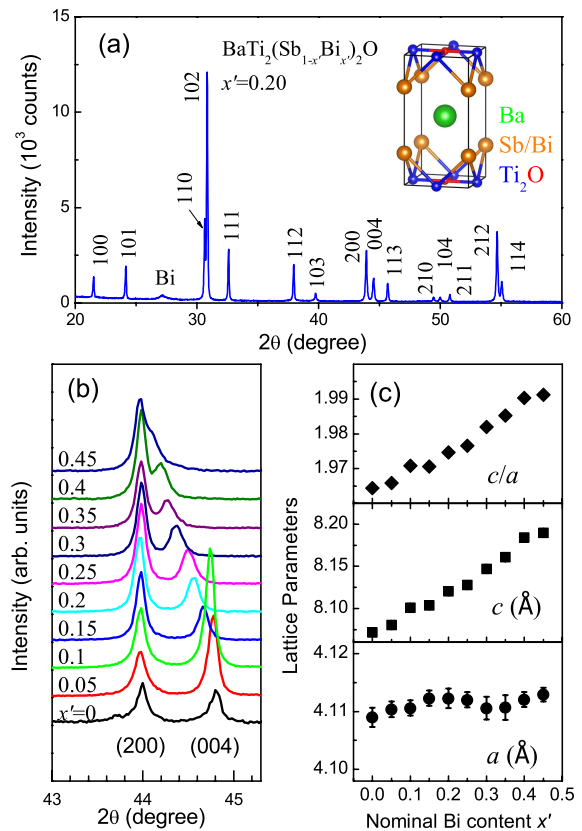}
\caption {(Color online) Structural characterizations of
BaTi$_{2}$(Sb$_{1-x'}$Bi$_{x'}$)$_{2}$O by x-ray diffraction with
CuK$_{\alpha1}$ radiation. (a) A typical XRD patterns of $x'$=0.20
sample indexed by the unit cell shown in the inset. (b) An expanded
plot showing systematic shift of (200) and (004) reflections with Bi
doping. (c) Lattice parameters as functions of nominal Bi content
$x'$.}
\end{figure}

With the Bi substitution, the (200) reflections move minutely, but
the (004) reflections shift remarkably, to the lower diffraction
angles, consistent with the mild increase in $a$-axis, yet
significant increase in $c$-axis [figure 1(c)]. The lattice
expansion indicates a negative chemical pressure by the Bi
substitution. Besides, the increase in $c/a$ ratio suggests
structural distortions in Ti$_2$Sb$_2$O-layers. It was found that
the fractional coordinate of each atom in the unit cell of
Na$_2$Ti$_2$Sb$_2$O did not change with the lattice constants upon
decreasing temperature.\cite{ozawa2000} This means that the $c/a$
ratio is basically proportional to the ratio of Sb-height to Ti-O
bondlength. The latter influences the crystal field splitting of the
Ti 3$d$ orbitals that was linked with the DW instability.\cite{TBA}
Since the $c/a$ ratio is significantly smaller for a DW
state,\cite{ozawa2000} and furthermore the $c/a$ value is only 1.8
for BaTi$_{2}$As$_{2}$O whose $T_{\text{DW}}$ is as high as 200
K,\cite{cxh1221} the increase in $c/a$ would be unfavorable for the
DW formation. As will be seen below, the DW ordering is indeed
suppressed by the Bi doping.

Figure 2 shows temperature dependence of resistivity [$\rho(T)$] of
the BaTi$_{2}$(Sb$_{1-x}$Bi$_{x}$)$_{2}$O polycrystalline samples.
The $\rho(T)$ data of the undoped compound resemble those of
previous reports\cite{yajima,doan} with regard to the DW anomaly at
$T_{\text{DW}}\sim$ 55 K, but the absolute resistivity is about one
order of magnitude smaller. The residual resistance ratio (RRR),
conventionally defined by the ratio of room-temperature resistance
and low-temperature one, is $\sim$ 20, much higher than those of
previous reports\cite{yajima,doan}. This suggests high quality of
the present samples. Upon Bi doping, $T_{\text{DW}}$ decreases
monotonically, and it vanishes for $x\geq$0.17. As stated above,
suppression of the DW state may be interpreted in terms of increase
of $c/a$ ratio associated with the negative chemical pressure.

\begin{figure}
\includegraphics[width=7cm]{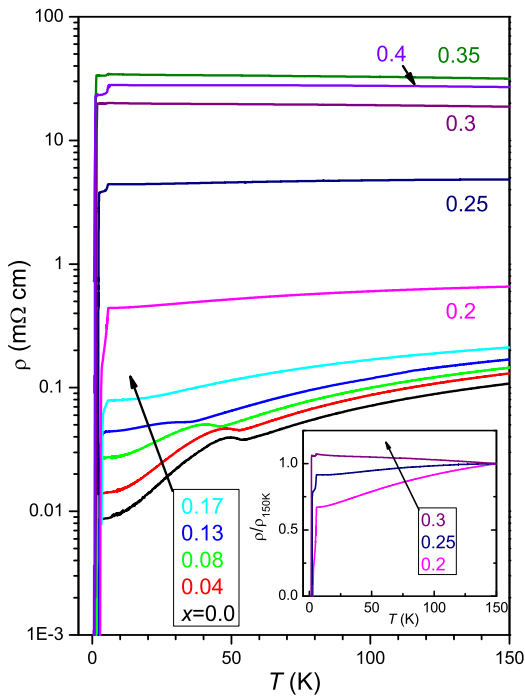}
\caption {(Color online) Temperature dependence of resistivity (in
logarithmic scale) for the  BaTi$_{2}$(Sb$_{1-x}$Bi$_{x}$)$_{2}$O
polycrystalline samples. The lower inset shows sign change in the
temperature coefficient of the normal-state resistivity around
$x$=0.3.}
\end{figure}

In addition to the negative chemical pressure effect, the Bi doping
induces disorder concomitantly. The low-temperature residual
resistivity increases almost proportionally with $x$ in the
low-doping regime. This can be explained by the conventional
impurity scattering due to the Sb/Bi substitution disorder. For the
high-doping samples, however, the absolute resistivity increases
rapidly. The low-$T$ resistivity increases by three orders of
magnitude from $x$=0 to $x$=0.3. Furthermore, the temperature
coefficient of resistivity (TCR) changes sign at $x\sim$0.3,
pointing to a metal-to-nonmetal (M-NM) transition. The origin of the
M-NM transition is probably a disorder-induced Anderson
localization,\cite{anderson} since the Bi substitution takes place
within the conducting Ti$_2$Sb$_2$O layers. Besides, the increase of
$c/a$ ratio reduces the dimensionality of the system, which may
aggravate the disorder effect.

\begin{figure}
\includegraphics[width=7cm]{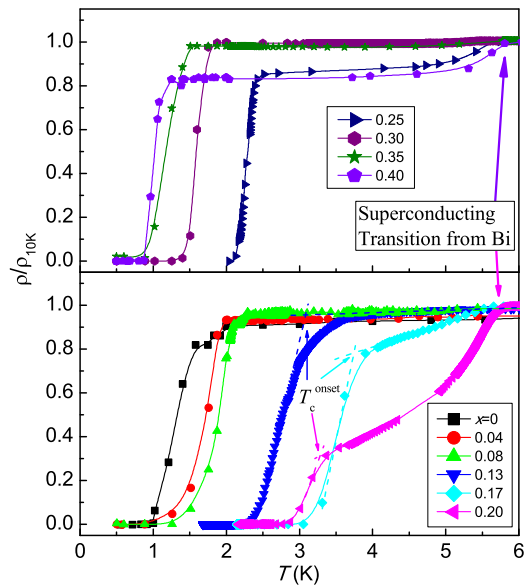}
\caption{(Color online) Superconducting transitions by the
resistivity measurement in BaTi$_{2}$(Sb$_{1-x}$Bi$_{x}$)$_{2}$O.
Definition of $T_c^{\text{onset}}$ is marked in the plot.}
\end{figure}

The detailed superconducting transitions in resistivity are
displayed in figure 3. The parent compound shows a superconducting
transition at $T_c^{\text{onset}}$=1.5 K, basically consistent with
the previous report.\cite{yajima} With the Bi doping, $T_c$ first
increases steadily up to 3.7 K at $x$=0.17, then decreases gradually
for $x\geq$ 0.2. A sharp superconducting transition at 0.8 K was
still seen for the highest-doping sample with $x$=0.40. Note that an
obvious resistivity drop always appears at 5.7 K for the high-doping
samples. These samples contain remarkable impurity of bismuth (about
5\% estimated from the relative intensities of the XRD peaks). This
impurity could become superconducting owing to the quenching-like
process during the sample preparations.\cite{note2}

Figure 4 shows temperature dependence of magnetic susceptibility
[$\chi(T)$] for the BaTi$_{2}$(Sb$_{1-x}$Bi$_{x}$)$_{2}$O samples
down to the lowest temperature available ($\sim$2 K). The magnetic
susceptibility under $H$=10 kOe indicates the DW anomaly (marked by
arrows) below 55 K for the low-doping samples. The anomaly is
weakened and, $T_{\text{DW}}$ decreases rapidly with the Bi doping.
No such anomaly can be detected for $x\geq$0.17. These results are
consistent with the above resistivity measurements shown in figure
2.

The superconducting diamagnetic transitions are also evident in the
temperature scope. For $x$=0.04 and 0.08 samples with lower $T_c$,
only superconducting onset transitions can be seen [the inset of
Fig. 4(d)]. The transition temperature determined by the magnetic
measurement ($T_{c}^{\text{mag}}$) agrees with the above resistivity
measurement. Note that the diamagnetic transition at 5.7 K for the
high-doping samples is again due to the superconducting transition
from the Bi impurity. The $\chi(T)$ data in the ZFC mode show
magnetic shielding fraction over 200\% (because the theoretical
density in the range of 6.0-6.7 g cm$^{-3}$ was employed, and no
demagnetization correction was made) for the samples with higher
$T_c$. The magnetic shielding signal is remarkably stronger than
those of the parent compound\cite{yajima} and Na-doped
BaTi$_{2}$Sb$_{2}$O\cite{doan}. In contrast, much lower diamagnetic
signal (less than 1\%) was measured in the FC mode. The vanishingly
small magnetic repulsion suggests strong magnetic-flux trapping,
which may be related to the Bi/Sb substitution disorder that could
serve as a flux-pinning center.

\begin{figure}
\includegraphics[width=8cm]{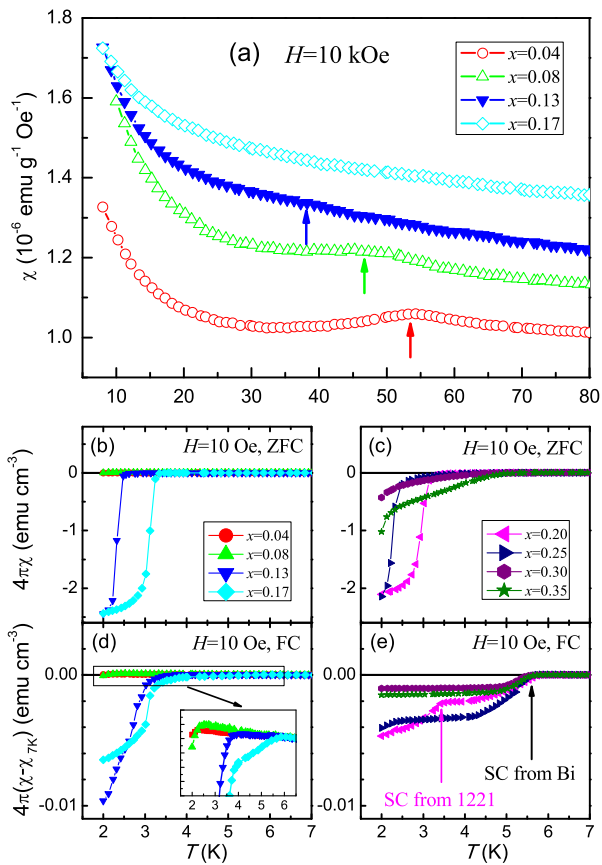}
\caption{(Color online) Temperature dependence of dc magnetic
susceptibility ($\chi$) for BaTi$_{2}$(Sb$_{1-x}$Bi$_{x}$)$_{2}$O.
The upper plot (a) and the lower panels [(b)-(e)] show high-field
and low-field data, respectively. Note that the field-cooling (FC)
data in panels (d) and (e) were subtracted by the $\chi$ value at 7
K for the clearness. Owing to the limit of the lowest temperature
available, only samples with $T_c>$ 2 K show superconducting
transitions.}
\end{figure}

Based on the above results, the electronic and superconducting phase
diagram can be established in figure 5. The parent compound
BaTi$_{2}$Sb$_{2}$O undergoes both a DW ordering at 55 K and a
superconducting transition at 1.5 K. The DW state is suppressed by
the Bi doping, and it disappears for $x\geq$0.17. Meanwhile the
superconducting transition temperature is elevated to its maximum
value $T_c$=3.7 K at $x$=0.17. A M-NM transition locates at $x\sim$
0.3. Nevertheless, SC does not vanish at the nonmetal side,
indicating that the SC is robust to disorder.

Very recently, Yajima \emph{et al.}\cite{note1} succeeded in
synthesizing the oxybismide end member BaTi$_{2}$Bi$_{2}$O, and an
enhanced $T_c$ of 4.6 K with no DW anomaly was reported. This result
is consistent with the case of lower doping ($x\leq$0.17)
BaTi$_{2}$(Sb$_{1-x}$Bi$_{x}$)$_{2}$O, which can be understood in
terms of the lattice expansion/distortion associated with negative
chemical pressure (see below). For the higher doping ($x\geq$0.2)
scenario, disorder effect induces Anderson localization and
suppresses $T_c$. Therefore, for the Bi-rich (0.5$<x<$1) regime, we
anticipate that $T_c$ would decrease monotonically if Bi is
partially substituted by Sb in BaTi$_{2}$Bi$_{2}$O.

Therefore, the Bi doping in the 1221 system brings about two
effects: lattice expansion and disorder. The former can be viewed as
a consequence of negative pressure, which also leads to structural
distortion of Ti$_{2}$Sb$_{2}$O-layers. The structural distortion of
Ti$_{2}$Sb$_{2}$O-layers, measured by the $c/a$ ratio, is closely
related to the DW state and possibly to SC also. The negative
chemical pressure increases the $c/a$ ratio, hence decreases
$T_{\text{DW}}$ and correspondingly increases $T_c$. However, the
maximum $T_c$ is still about1 K and 2 K lower than those of
BaTi$_{2}$Bi$_{2}$O\cite{note1} and the Na-doped
BaTi$_{2}$Sb$_{2}$O\cite{doan}. This implies that the Bi/Sb
substitution disorder plays a role. In the high-doping region, such
disorder leads to an Anderson localization that is responsible for
the observed M-NM transition. Surprisingly, SC still appears when
the normal state shows nonmetallic behaviors (negative TCR and
relatively high resistivity). This fact suggests a possible
realization of "fractal superconductivity" near the localization
threshold in the present system.\cite{fractal sc} This interesting
issue deserves future explorations.

\begin{figure}
\includegraphics[width=7cm]{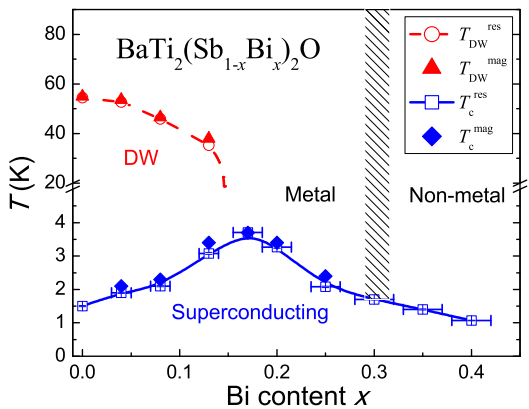}
\caption{(Color online) Electronic and superconducting phase diagram
of BaTi$_{2}$(Sb$_{1-x}$Bi$_{x}$)$_{2}$O. DW refers to spin- or
charge-density wave state. The phase boundaries are determined by
both resistivity ($T^{\text{res}}$) and magnetic susceptibility
($T^{\text{mag}}$) measurements. Note that the vertical axis was
broken from 5 to 20 K in order to show the related phases clearly.}
\end{figure}

In summary, we have investigated an isovalent substitution effect on
SC and CDW/SDW in BaTi$_{2}$(Sb$_{1-x}$Bi$_{x}$)$_{2}$O system. The
Bi doping induces negative chemical pressure which not only expands
but also distorts the lattice. It was found that the CDW/SDW
ordering was completely suppressed at $x=$0.17, and concomitantly
the superconducting transition temperature was elevated to the
maximum $T_c$ of 3.7 K. A metal-to-nonmetal transition at $x\sim$
0.3 was observed, which is interpreted by Anderson localization due
to the Bi/Sb substitution disorder. Interestingly, such disorder
does not kill the superconductivity. These results supply some
useful clues to further study the nature of the CDW/SDW phase and
its relations to superconductivity in BaTi$_{2}$Sb$_{2}$O-related
systems.

\begin{acknowledgments}
This work is supported by the NSF of China (No. 11190023), the
National Basic Research Program of China (Nos. 2010CB923003 and
2011CBA00103), and the Fundamental Research Funds for the Central
Universities of China.
\end{acknowledgments}

\end{document}